%% file: main.tex
\documentclass[%
prl,
twocolumn,
superscriptaddress,
 amsmath,amssymb,
 aps,
floatfix,
]{revtex4-2}

\usepackage[utf8]{inputenc}

\usepackage{graphicx}% Include figure files
\usepackage{dcolumn}% Align table columns on decimal point
\usepackage{bm}% bold math
\usepackage{isotope}
\usepackage{upgreek}  % upright greek letters
\usepackage{multirow}  % multirow in table
\usepackage{xcolor}
\usepackage{placeins}           % \FloatBarrier
\usepackage[colorlinks=true,citecolor=blue,filecolor=blue,linkcolor=blue,urlcolor=blue]{hyperref}
\usepackage{cleveref}
\usepackage[separate-uncertainty=true,per-mode=symbol]{siunitx} %use for _all_ units
\usepackage[caption=false]{subfig}
\usepackage{lipsum} %dummy text for layout
\usepackage{comment}
\usepackage[normalem]{ulem}

\DeclareSIUnit\clight{\text{\ensuremath{c}}}
\DeclareSIUnit\tevm{\TeV\per\clight\squared}
\DeclareSIUnit\events{events}
\DeclareSIUnit\years{y}
\DeclareSIUnit\tonne{t}
\DeclareSIUnit\tonneyears{\tonne\years}
\DeclareSIUnit\cmtwo{\cm\squared}
\DeclareSIUnit\kev{\kilo\eV} 
\DeclareSIUnit\mev{\mega\eV} 
\DeclareSIUnit{\GeV}{\giga\eV}
\DeclareSIUnit{\TeV}{\tera\eV}
\DeclareSIUnit{\gevm}{\GeV\per\clight\squared}
\DeclareSIUnit\kevnr{\si{\kev} {}_\mathrm{NR}} 
\DeclareSIUnit\kever{\si{\kev} {}_\mathrm{ER}} 
\DeclareSIUnit\PE{\mathrm{PE}}
\DeclareSIUnit\bar{bar}

\newcommand{\nc}[2]{\newcommand{#1}{#2}#2}

\newcommand{\radius}{\ensuremath{\mathrm{R}}}
\newcommand{\depth}{\ensuremath{\mathrm{Z}}}
\newcommand{\rnttz}{\isotope[220]{Rn}}
\newcommand{\argon}{\isotope[37]{Ar}}
\newcommand{\kretm}{\isotope[83\textrm{m}]{Kr}}
\newcommand{\ambe}{\isotope[241]{AmBe}}
\newcommand{\er}{ER}
\newcommand{\nr}{NR}
\newcommand{\cevns}{\ensuremath{\mathrm{CE}\nu\mathrm{NS}}}

\newcommand{\tapprox}{\raisebox{0.5ex}{\texttildelow}}

\newcommand{\qmark}[1]{``#1''} %quotation mark

\newcommand{\fiducialvolume         }{\SI{4.18}{\tonne}}%https://xe1t-wiki.lngs.infn.it/doku.php?id=xenon:xenonnt:analysis:ntsciencerun0:fiducial_volume
\newcommand{\fiducialvolumeincludingerror}{\SI{4.18\pm0.13  }{\tonne}}
\newcommand{\uncorrectedlivetime               }{\SI{97.1}{\day}}%https://xe1t-wiki.lngs.infn.it/doku.php?id=xenon:xenonnt:analysis:ntsciencerun0:wimp_unblinding_proposal
\newcommand{\livetime               }{\SI{95.1}{\day}}%https://xe1t-wiki.lngs.infn.it/doku.php?id=xenon:xenonnt:analysis:ntsciencerun0:wimp_unblinding_proposal
\newcommand{\exposure               }{\SI{1.09 \pm 0.03}{\tonneyears}}%TODO fill value

\newcommand{\tpctemperaturesrzero}{\ensuremath{(176.8\pm 0.4)}\,\si{\kelvin} } %private_nt_aux
\newcommand{\bellpressuresrzero}{\ensuremath{(1.890\pm0.004)}\,\si{\bar}} %lowER value (should check with Shingo)
\newcommand{\elife}{\ensuremath{\SI{10}{\milli\s}}} %lowER numbers
\newcommand{\purificationflow}{\ensuremath{\SI{8.3}{\tonne\per\day}}}
\newcommand{\totalxenonmass}{\SI{8.5}{\tonne}} %lowER paper
\newcommand{\activexenonmass}{\SI{5.9}{\tonne}}

\newcommand{\excludedpmts           }{\num{17}}%https://xe1t-wiki.lngs.infn.it/doku.php?id=xenon:xenonnt:dsg:pmt:gains:time_dependent_model#the_current_status_of_the_gain_model
\newcommand{\averagepmtgain }{\num{2e6}}%https://xe1t-wiki.lngs.infn.it/doku.php?id=xenon:xenonnt:dsg:pmt:gains:time_dependent_model
\newcommand{\pmtgainstability  }{\SI{3}{\percent}}%https://xe1t-wiki.lngs.infn.it/doku.php?id=xenon:xenonnt:dsg:pmt:gains:time_dependent_model

\newcommand{\driftfield             }{\SI{23}{V \per cm}}

\newcommand{\wirecutwidth}{\SI{8.9}{\cm}} %from https://github.com/XENONnT/cutax/blob/ac33dfead6d499f59ee918bcb251d2aed3085772/cutax/cuts/fiducial_volume.py#L82
\newcommand{\errateperkev}{\SI{15.8\pm1.3}{\events\per\tonne\per\years\per\kev}} %from lowER paper
\newcommand{\ambebranching}{\SI{60}{\percent}}
\newcommand{\ambegammaenergy}{\SI{4.44}{\mev}}

\newcommand{\nvetotaggingefficiency}{\SI{53\pm3}{\percent}}

\newcommand{\nrbackgroundevents}{$1.1^{+0.6}_{-0.5}$}

\newcommand{\cesw}{\SI{13.7}{\electronvolt\per\mathrm{quantum}}}% https://xe1t-wiki.lngs.infn.it/doku.php?id=xenon:xenon1t:hoetzsch:bbf:bbf_nrfitsr0:summary
\newcommand{\cesgp}{\SI{0.151(1)}{\PE\per\mathrm{photon}}}
\newcommand{\cesge}{\SI{16.5(6)}{\PE\per\mathrm{electron}}}

\newcommand{\sonerangelow}{0}
\newcommand{\sonerangehigh}{100}

\newcommand{\stworangelow}{126}
\newcommand{\stworangehigh}{12589}

\newcommand{\energyrangenrlow}{\SI{3.3}{\kevnr}}%TODO fill value
\newcommand{\energyrangenrhigh}{\SI{60.5}{\kevnr}}%TODO fill value
\newcommand{\energyrangenrlownonr}{\SI{3.3}{\kev}}%TODO fill value
\newcommand{\energyrangenrhighnonr}{\SI{60.5}{\kev}}%TODO fill value
\newcommand{\energyrangenr}{\energyrangenrlow-\energyrangenrhigh}

\newcommand{\sciencebinnedgof}{\num{0.63}} %https://xe1t-wiki.lngs.infn.it/doku.php?id=xenon:xenonnt:zihao:bbf_sr0_er_fitting_update best-fit at 200GeV

\newcommand{\wimpmassbest           }{\SI{28}{\gevm}}
\newcommand{\limitmassbest}{\SI{2.58e-47}{\cmtwo}}% 13.4
\newcommand{\highmasslimitfunction}{$\SI{6.08e-47}{\cmtwo}\times(M_\mathrm{DM} /( \SI{100}{\gevm}))$} %13.4

\newcommand{\minimumpvalue}{\ensuremath{0.20}} %v13.4

\newcommand{\bestfitfreesignalxsec }{\SI{3.22e-47}{\cmtwo}} %13.5
\input{expectation_definitions}

\newcommand{\nominaler}{$134$}
\newcommand{\nominalradiogenictotal}{\nrbackgroundevents} %byhand Also note that in the inference, these constraints are symmetrized

\newcommand{\neventstotal}{$152$}
\newcommand{\neventsblinded}{$16$}
\newcommand{\neventswire}{$28$}
\newcommand{\neventsnowire}{$124$}

\newcommand{\neventsupperrighthalf}{$13$} %blinded events with 0<x+y 

\newcommand{\neventsreferenceregion}{$3$}
\newcommand{\neventsreferenceregionwire}{$1$}
\newcommand{\neventsreferenceregionnonwire}{$2$}

\newcommand{\neventsrncalibration}{$2051$}%in the region of interest 
\newcommand{\rnttzpval}{\ensuremath{0.18}}
\newcommand{\ambepval}{\ensuremath{0.39}}

\begin{document}
% =====================================

% =====================================
% Make title
% =====================================
\title{First Dark Matter Search with Nuclear Recoils from the XENONnT Experiment}

\include{affiliations}
\include{author}
\collaboration{XENON Collaboration}
\email[]{xenon@lngs.infn.it}
\noaffiliation

\date{\today}

\begin{abstract}
\newpage
We report on the first search for nuclear recoils from dark matter in the form of weakly interacting massive particles (WIMPs) with the XENONnT experiment which is based on a two-phase time projection chamber with a sensitive liquid xenon mass of {\activexenonmass}. During the \exposure~exposure used for this search, the intrinsic \isotope[85]{Kr} and \isotope[222]{Rn} concentrations in the liquid target were reduced to unprecedentedly low levels, giving an electronic recoil background rate of {\errateperkev} in the region of interest.
A blind analysis of nuclear recoil events with energies between {\energyrangenrlownonr} and {\energyrangenrhighnonr} finds no significant excess. This leads to a minimum upper limit on the spin-independent WIMP-nucleon cross section of {\limitmassbest} for a WIMP mass of {\wimpmassbest} at \SI{90}{\percent} confidence level. Limits for spin-dependent interactions are also provided. Both the limit and the sensitivity for the full range of WIMP masses analyzed here improve on previous results obtained with the XENON1T experiment for the same exposure. 
\end{abstract}

\maketitle

% =====================================
% Main content
% =====================================
Astrophysical and cosmological observations indicate the existence of a massive, non-luminous, non-relativistic and non-baryonic dark matter (DM) component of the Universe~\cite{Bertone:2004pz}. One well-motivated class of DM candidates is weakly interacting massive particles (WIMPs), which arise naturally in several beyond-Standard-Model theories~\cite{Roszkowski:2017nbc}. 
Direct detection searches for WIMPs with masses of a few \si{\gevm} to tens of \si{\tevm} using liquid xenon (LXe) time projection chambers (TPCs) have produced the most stringent limits to date on elastic spin-independent WIMP-nucleon cross sections~\cite{XENON:2018voc,PandaX:2021pan,LZ:2022dm}.

The XENON Dark Matter project currently operates the XENONnT experiment at the INFN Laboratori Nazionali del Gran Sasso (LNGS) underground laboratory. It is an upgrade of its predecessor, XENON1T~\cite{XENON:2017lvq}, with a new, larger dual-phase TPC featuring a sensitive LXe mass of {\activexenonmass}. The XENON1T cryogenics, gaseous purification and krypton distillation systems, as well as the \SI{700}{\tonne} water Cherenkov muon veto (MV) tank~\cite{XENON1T:2014eqx, XENON:2020sensi} are reused to operate XENONnT. 
Inside the water tank, a new neutron veto (NV) detector encloses the TPC cryostat.
For the exposure used in this analysis, the NV was operated as a water Cherenkov detector, tagging neutrons through their capture on hydrogen which releases a \SI{2.22}{\mega\electronvolt} $\upgamma$-ray.

The senstive LXe detector volume, enclosed by a polytetrafluoroethylene (PTFE) cylinder with a height of \SI{1.49}{\meter} and a diameter of \SI{1.33}{\meter},
is viewed by 494 Hamamatsu\,R11410-21 \mbox{3-inch} photomultiplier tubes (PMTs)~\cite{Antochi:2021pmt} distributed in a top and a bottom array.
To fill the vessel housing the TPC, a total of {\totalxenonmass} liquified xenon is required which is continuously purified by a new liquid-phase purification system~\cite{Plante:2022khm}. Together with a high flow radon distillation system~\cite{Murra:2022mlr}, a careful selection of detector construction materials~\cite{XENON:2021mrg} and a specialized assembly procedure, this led to an unprecedentedly low electronic recoil (\er) background of {\errateperkev} below recoil energies of \SI{30}{\kev}~\cite{XENON:2022lower}.

Particles depositing energy in the LXe produce a prompt scintillation signal (S1) as well as ionization electrons which drift upwards and are extracted into the gas above the liquid due to applied electric fields. 
Here a second scintillation signal (S2), proportional to the number of extracted electrons, is produced. 
WIMPs are expected to primarily produce nuclear recoils ({\nr}s), where a xenon nucleus recoils, while the background is dominated by {\er} interactions where an electron recoils. 
A higher scintillation-to-ionisation ratio is expected for {\nr}s, but unlike {\er}s, a fraction of the total recoil energy is also lost as unobservable heat.

Three parallel-wire electrodes (cathode, gate and anode) are used to establish the drift and extraction fields. The gate and anode electrodes are reinforced with two and four transverse wires, respectively, to minimize wire sagging.
Two additional parallel-wire screening electrodes are used to shield the PMT arrays from the electric fields. 
After two months of commissioning at a drift field of $\tapprox$\SI{100}{\volt\per\cm}, a short between the bottom screening and cathode electrodes limited the applied drift field to {\driftfield}, corresponding to a maximum drift time of \SI{2.2}{\milli \second}.
The extraction field was set to \SI{2.9}{\kilo\volt\per\cm} in LXe to reduce localized, intermittent bursts of single electron S2 signals.
Despite the lower-than-designed drift and extraction fields, the energy and position resolution, as well as the energy threshold, are comparable to those achieved with XENON1T.

The TPC and veto detectors are integrated into a single data aquisition system~\cite{XENON:2023daq}. The data acquired by the MV uses the same hardware event trigger as in XENON1T~\cite{XENON:2019bth}, whereas data from the TPC and NV are acquired in a \qmark{triggerless} mode, with each individual PMT channel recording all signals above a channel-specific threshold of $\tapprox\SI{0.13}{}$ photoelectrons (\si{\PE}). %from Joran https://xe1t-wiki.lngs.infn.it/doku.php?id=xenon:xenon1t:org:papers:nt:sr0-wimp:nikhef_v10

The recorded signals are processed using custom-developed open source software packages~\cite{strax,straxen}. Each PMT signal is scanned for PMT \qmark{hits} above threshold, and hits found in the TPC channels are clustered and classified into S1, S2 or \qmark{unclassified} peaks based on pulse shape and PMT hit-pattern. At least three PMTs must contribute to an S1 within \SI{\pm 50}{\nano\second} around the center of the integrated peak waveform. Events are built in time intervals between \SI{2.45}{\milli \second} before and \SI{0.25}{\milli \second} after S2s, and overlapping events are merged. The event S2 is required to be greater than \SI{100}{\PE}, and have fewer than eight other peaks larger than half of the S2 peak area within \SI{\pm 10}{\ms}. 

The PMT hit patterns of S2 signals are used to reconstruct the horizontal position ($\mathrm{X},\mathrm{Y}$) of an event using neural network models~\cite{XENON:2019ykp,MLP_Shixiao}. Each model was trained by the S2 light distribution on the top PMT array generated through optical simulations with Geant4 \cite{XENON:2020sensi}, corrected for the number of exlcuded PMTs and electronics per-PMT response with the XENONnT waveform simulator (WFSim)~\cite{wfsim}.  
The horizontal interaction position resolution for simulated events close to the PTFE detector walls is \tapprox$\SI{1}{\centi\meter}$, and \tapprox$\SI{0.75}{\centi\meter}$ within the fiducial volume (FV), for a \SI{1000}{PE} S2 (\tapprox\SI{30}{extracted\ electrons}).
The depth, \depth, of an interaction is reconstructed from the measured drift time between S1 and S2 and the electron drift velocity with a resolution $<\SI{1}{\%}$. 
The \SI{50}{\percent} S2 width of a single electron signal is about \SI{600}{\nano \second} and the width of S2s within the FV of the detector typically range from \SI{2}{\micro \second} to \SI{9}{\micro \second}.
The drift field has a radial component that shifts ionization electrons originating deeper in the detector inwards when they are observed at the liquid surface.
This inward shift is corrected with a data-driven approach, assuming a uniform distribution of {\kretm} calibration events in radius squared ({\radius}$^2$) as in~\cite{XENON:2019ykp}. 

The position and time information of the detected S1 and S2 signals is used to correct for the inhomogeneous detector response due to quanta generation and collection effects, and corresponds to corrections of up to \SI{30}{\%} for either signal. 
Scintillation photons are affected by a position-dependent optical light collection efficiency which reduces the S1 peak area. 
A light yield (LY) map normalized to the mean response in the (FV) is generated using {\kretm} signals. 
The electric field dependence of the LY is removed using a drift field map constructed by matching the spatial distribution of {\kretm} to a COMSOL~\cite{comsol} simulation, accounting for potential charge accumulations on the PTFE surfaces. 
This drift field map was validated with data using the measured S1 ratio of the two {\kretm} decays~\cite{LUX:2017kr}. 
The resulting LY map is valid over the full energy range of this analysis and is used to correct S1 signals, referred to as cS1.

The S2 peak area reduces exponentially for signals deeper in the detector, as drifting electrons can be captured by electronegative impurities. This effect leads to a time-dependent lifetime of the free electrons which is corrected using data from {\kretm} and \isotope[222]{Rn} decays, and monitored with a new purity monitor system~\cite{icarus:pm}.  
The charge yield of the respective sources was corrected by the drift field map using low-field data from~\cite{Jorg:2021hzu}. An electron lifetime better than {\elife}  was reached throughout the science run with a liquid purification flow of {\purificationflow}~\cite{Plante:2022khm}.
The spatial variation in the S2 response is dominated by the position-dependent optical light collection efficiency and inhomogeneous electroluminescence amplification. {\kretm} events are used to obtain a normalized horizontal S2 peak area correction map. 
Time-dependent variations of the single electron gain and extraction efficiency following each ramping up of the electric field are corrected by their respective data-driven trends. 
S2 signals summed over the top and bottom array, and corrected for the above effects are referred to as cS2. 

The method to convert the cS1 and cS2 signals of {\nr}s and {\er}s into a combined energy scale is described in~\cite{XENON:2020rca}. 
The photon and electron gains are found to be $g_1 = \cesgp$ and $g_2 = \cesge$, assuming the mean energy to produce a charge or light quantum to {\cesw} \cite{Dahl_w_value}. Reconstructed energies using this scale directly give the {\er}-equivalent energy (\si{\kever}), while the {\nr}-equivalent energy (\si{\kevnr}) requires a model for energy lost to heat, and uses the full {\nr} detector model, described later. 

The science search data was collected from \nc{\srzerosciencestart}{July 6$^\text{th}$}~to~ \nc{\srzeroscienceend}{November 10$^\text{th}$ 2021}. This period, named Science Run 0 (SR0), contains a total of {\uncorrectedlivetime} of data which corresponds to a deadtime- and veto-corrected livetime of {\livetime}.
The length of SR0 was primarily chosen to investigate the XENON1T {\er} excess~\cite{XENON:2020rca}, leading to a WIMP search exposure of {\exposure}. 
The detector conditions were stable throughout SR0 with an average LXe temperature of {\tpctemperaturesrzero} and  pressure of {\bellpressuresrzero}, where the uncertainties represent the corresponding RMS over SR0. PMT gains were monitored by weekly calibrations with a pulsed low-intensity light source and voltages were adjusted at the beginning of SR0 to achieve \tapprox$\averagepmtgain$ gains for all PMTs. The time dependence of the PMT gains was modeled and the signals were corrected, resulting in a gain variation $<{\pmtgainstability}$. In total {\excludedpmts} PMTs were excluded from analysis due to internal vacuum degradation, instability, light emission or noise. Five of these PMTs are distributed evenly in the top PMT array.
Periods of data taken with an intermittent and localized high rate of S2 emission from single or few electrons are not included in calibration and search data. 
Calibrations with {\kretm} were performed every second week to correct the detector response for position- and time-dependent effects, and to monitor the stability of cS1 and cS2.

\begin{figure}[t]
    \centering
    \includegraphics[width=\columnwidth]{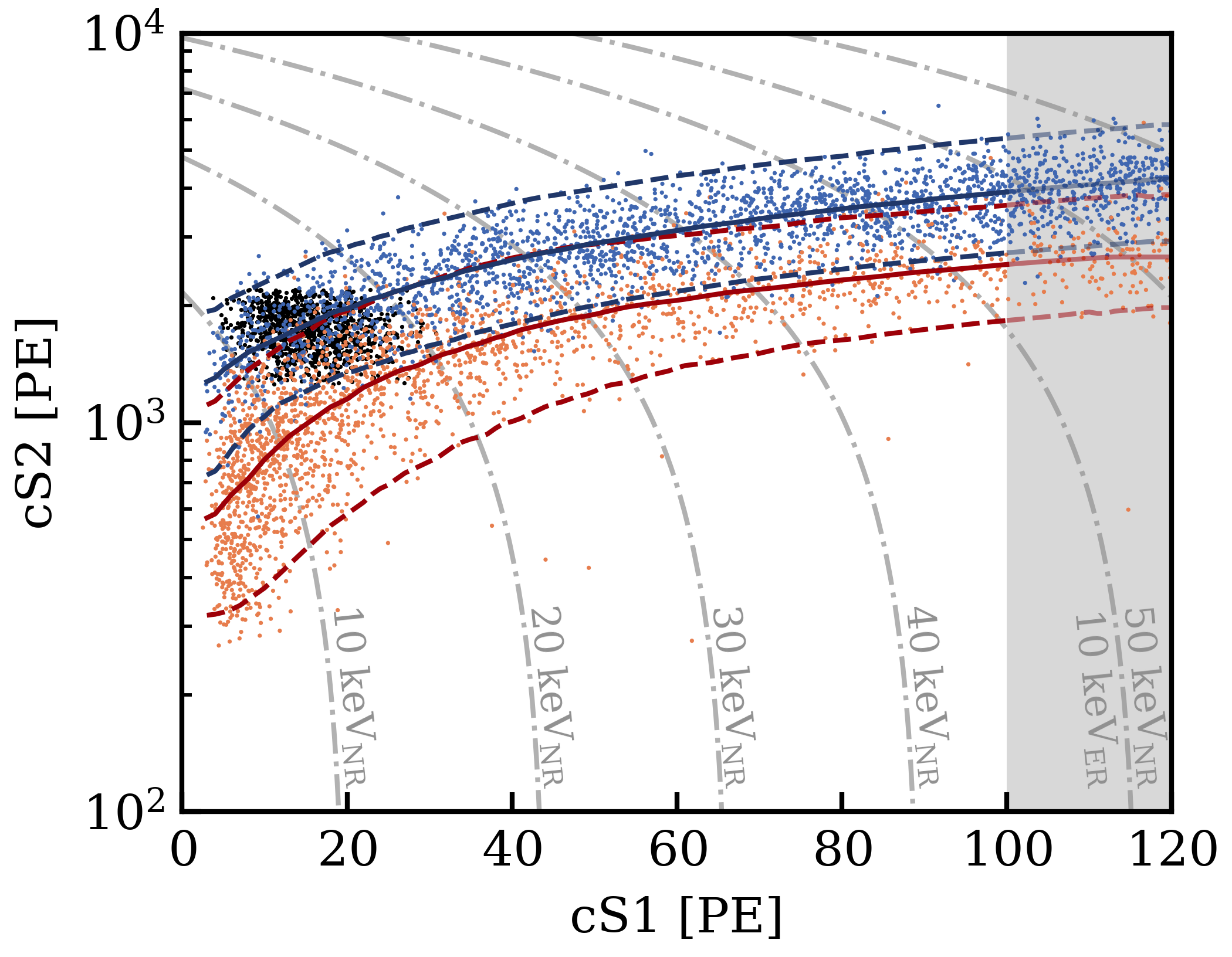}
    \caption{{\nr} and {\er} calibration data from {\ambe} (orange),  \isotope[220]{Rn} (blue) and {\argon} (black). The median and the $\SI{\pm2}{\sigma}$ contours of the NR and ER model are shown in blue and red respectively. The gray dash-dotted contour lines show the reconstructed {\nr} energy (\si{\kevnr}). Only not shaded events up to a cS1 of \SI{100}{\PE} are considered in the response model fits.}
    \label{fig:nr_er_calibration_data}
\end{figure}

The {\nr} response of XENONnT and the NV tagging efficiency were calibrated using an external {\ambe} source which was placed in three positions close to the TPC cryostat. {\ambe} emits neutrons via the alpha-capture reaction $\isotope[9]{Be}(\upalpha, n)\isotope[12]{C}$ which has a chance of about {\ambebranching} to emit an additional {\ambegammaenergy} $\upgamma$-ray~\cite{ambe_scherzinger}. This $\upgamma$-ray, well above the NV threshold, is used to select {\nr} S1 signals in a $\tapprox$\SI{400}{\nano \second} window. After applying the same data-quality cuts as used in the main analysis, 1986 events remain in the region of interest (ROI), shown in Figure~\ref{fig:nr_er_calibration_data}. Only \SI{1.8(6)}{} events are expected from random coincidences between the two detectors, determined through a sideband study. 
The tagging efficiency of the NV is estimated from the number of delayed neutron capture signals following the NR S1 signals. 
This data-driven tagging efficiency is corrected for position-dependent effects using Geant4~\cite{GEANT4:2002zbu} simulations which account for the full spatial distribution of neutrons emitted by detector materials~\cite{XENON:2020sensi}. 
The length of the veto window was set to \SI{250}{\micro \second} with a 5-fold PMT coincidence and a \SI{5}{\PE} event area threshold in the NV. This gives a neutron tagging efficiency of {\nvetotaggingefficiency}, and a livetime reduction of \SI{1.6}{\percent}.

The {\er} response model is calibrated with {\neventsrncalibration} \isotope[212]{Pb} $\upbeta$ events from a \isotope[220]{Rn} calibration source~\cite{XENON:2016rze}, before SR0 and with events from an \isotope[37]{Ar} source~\cite{XENON:2022ar37} collected after SR0, as discussed in~\cite{XENON:2022lower}. 
{\nr} and {\er} calibration datasets were fitted using the LXe response model and fast detector simulation described in~\cite{xe1t_ana_paper_2}. 
For both datasets, a Markov-Chain Monte Carlo (MCMC) sampling of the parameter space gives the best-fit point and posterior distribution. The goodness-of-fit (GOF) was assessed by partitioning the cS1, cS2 space into equiprobable bins according to both best-fit models and then computing a Poisson $\chi^2$ likelihood, as well as one-dimensional projections on cS2. 
Neither tests reject the best-fit model, with two-dimensional p-values of {\rnttzpval} and {\ambepval} for {\er} and {\nr}, respectively, and no significant p-values for the one-dimensional projections.
The calibration data and contours of the best-fit model are shown in Figure \ref{fig:nr_er_calibration_data}. The leakage fraction of the \isotope[220]{Rn} ER events below the NR median is $1.1^{+0.2}_{-0.3}\,\%$.

The full {\er} model has too many parameters to be tractable in the inference toy MC simulations. Using linear combinations of the original parameters identified with a principal component analysis reduces parameter redundancies, and these parameter directions are then ranked according to their impact on the background expectation in a signal-like region in cS1 and cS2.
The two parameters with the highest impact are included as nuisance parameters in the {\er} model used in the WIMP search likelihood.
 
The ROI is defined by cS1 between \SI{\sonerangelow}{\PE} and \SI{\sonerangehigh}{\PE} and cS2 between \SI{\stworangelow}{\PE} and \SI{\stworangehigh}{\PE}. Together with detection and selection efficiencies, this gives an energy range with at least \SI{10}{\percent} total efficiency from {\energyrangenr}.
All events reconstructed with an {\er} energy below \SI{20}{\kever} and found in the cS1 and cS2 contours of the {\er} and {\nr} band were blinded. For the study of the {\er} data presented in~\cite{XENON:2022lower} all events above the $\SI{-2}{\sigma}$ quantile of the {\er} band or with a reconstructed {\er} energy larger than \SI{10}{\kever} were unblinded. The remaining region was unblinded only after finalizing the analysis procedure presented here.

The event selection criteria from~\cite{XENON:2019ykp} were optimized for the ROI in this analysis. Data quality cuts are applied in order to include only well-reconstructed events and to suppress backgrounds. All cuts were optimized based on calibration data and simulations using WFSim. Each valid event is required to have a valid S1-S2 pair. Events tagged by the MV or NV are removed from the data selection as are multiple-scatter (MS) events since WIMPs are expected to induce only single-scatter (SS) {\nr}s.
The MV uses a veto window of \SI{1}{\milli \second} with a 5-fold PMT coincidence and a \SI{10}{\PE} MV event area threshold.

A dedicated cut similar to that in~\cite{XENON:2020gfr} using a gradient boosted decision tree (GBDT) was developed to reduce the background due to randomly paired S1-S2 signals called accidental coincidences (ACs).
This cut uses S2 area and shape, as well as interaction depth, and reduces the AC background by \SI{65}{\%} at \SI{95}{\%} signal acceptance. 
Due to an insufficient model of the S2 pulse shape near the transverse wires, caused by local variations of the drift and extraction field with respect to the rest of the TPC, an optimization of the GBDT and other S2 shape-based cuts was not possible with WFSim.
Consequently, the LXe target is split into two parts in the modeling for the WIMP search. A less strict data-driven model for the S2 width cut and no GBDT selection is used in an {\wirecutwidth} wide band around the transverse wires, leading to a lower signal-to-background ratio, but with a \SI{10}{\percent} higher selection efficiency. The total selection efficiency for these \qmark{near}- and \qmark{far}-wire regions is estimated following the procedure in~\cite{XENON:2020rca,XENON:2019ykp}. Efficiency losses due to the event building are also taken into account in the selection efficiency.

\begin{figure}[t]
    \centering
    \includegraphics[width=\columnwidth]{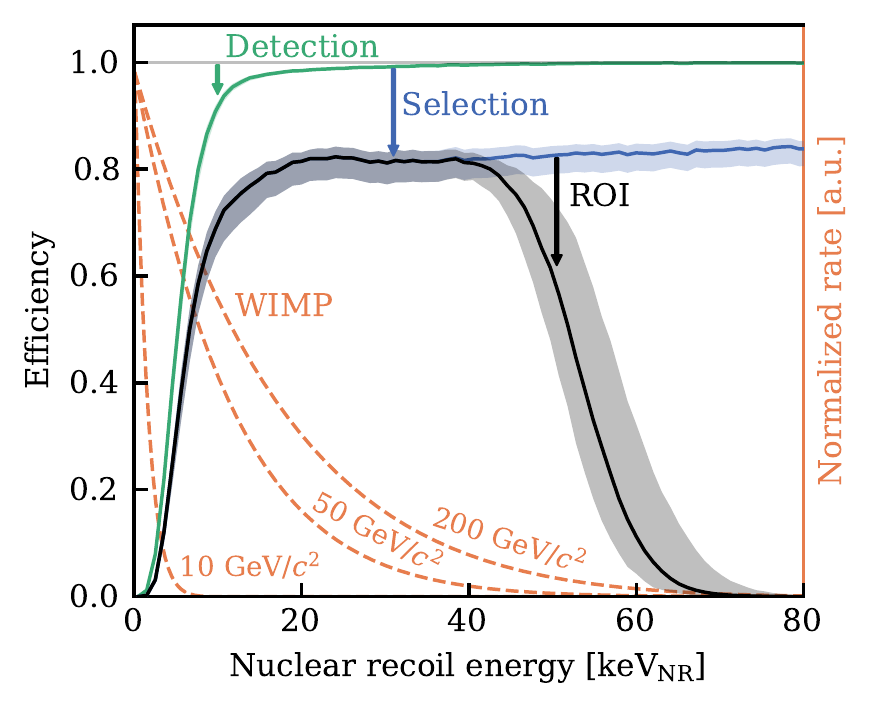}
    \caption{Detection and selection efficiency for {\nr} events in this search as a function of the {\nr} recoil energy. The total efficiency in the WIMP search region (black) is dominated by the detection efficiency (green) at low energies and event selections (blue) at higher energies until the edge of the ROI. 
    Normalized recoil spectra for WIMPs with masses of \SI{10}{\gevm}, \SI{50}{\gevm}, and \SI{200}{\gevm} are shown with orange dashed lines for reference.}
    \label{fig:efficiency_and_acceptance}
\end{figure}

The detection efficiency of the TPC, dominated by the S1 detection efficiency, is evaluated using WFSim and validated with a data-driven method~\cite{xe1t_ana_paper_2,pema}. Both methods agree within \SI{1}{\percent}. Efficiency losses at small energies are dominated by the 3-fold PMT coincidence requirement. The upper cS1 ROI edge, chosen to include the full WIMP spectrum, determines the upper edge of this analysis.
The combined selection efficiency of the near and far wire regions, the detection and the total efficiencies of the analysis are shown together with the normalized recoil spectra of three different WIMP masses in Figure~\ref{fig:efficiency_and_acceptance}. 

% detection efficiency

In order to mitigate background events from detector radioactivity as well as \qmark{surface events} produced by {\er}s from \isotope[210]{Pb} plate-out~\cite{XENON:2018voc}, only events reconstructed in a central FV (illustrated in Figure~\ref{fig:data_rz} in the supplementary material) are considered in the analysis. 
The FV shape is optimised based on the background distributions, as well as constrained to not include regions where the detector is not sensitive or models are incomplete.
The total LXe mass of the FV after considering the systematic uncertainty of the field distortion correction is {\fiducialvolumeincludingerror}. 

Five different background components make up the total background model: radiogenic neutrons, coherent elastic neutrino-nucleus scattering ({\cevns}), {\er}s, surface events and ACs. The expectation values for each are summarized in Table \ref{tab:backgrounds}. In addition to the full expectation values, we include for illustration expectation values in a signal-like region defined to contain half of a \SI{200}{\gevm} WIMP signal with the lowest signal-to-background ratio.

The {\nr} background in XENONnT is dominated by radiogenic neutrons from spontaneous fission and ($\upalpha$,~$n$)-reactions. Neutron yields and energies originating from various detector materials are evaluated as in~\cite{xe1t_ana_paper_2, XENON:2020sensi}.
A custom interface based on the fitted {\nr} model accepts Geant4  simulation inputs, and provides observable quanta processed by WFSim to construct the neutron background model~\cite{EPIX}.
The neutron rate was estimated based on this full detector simulation and compared against a data-driven method. 
The data-driven estimate uses a combined Poisson likelihood for MS and SS events tagged by the NV, together with a simulation-driven MS/SS ratio which was validated with {\ambe} data. The maximum deviation of the MS/SS ratio estimated as a function of radius between data and simulation was found to be less than \SI{20}{\percent}.
However, a wrong sign in the NV tagging window, discovered only after unblinding of the main data, meant that the simulation and data-driven estimates found before were no longer in agreement. This error arose from the premise that the tagging efficiency was determined in a forward coincidence, counting the number NV tags for a given set of NR SS events, while the tagging is done by a backwards veto triggered when a NV event satisfies the threshold criteria. In accordance with the analysis plan, the data-driven rate estimate is used. 
Four events in the WIMP blinding region are tagged by the NV and cut, three of them also fail the SS cut, compatible with the MS/SS ratio from simulations. This gives a total neutron expectation of {\nominalradiogenictotal} events which is a factor \tapprox\SI{6}{} higher than predicted by simulations. Analysis choices such as the NV tagging window and the FV were not re-optimized after this correction.
\setlength\extrarowheight{2pt}
\begin{table}[t]
    \centering
    \caption{
    Expected number of events for each model component and observed events. 
    The \qmark{nominal} column shows expectation values and uncertainties, if applicable, before unblinding. 
    The nominal ER value is the observed number of ER events before unblinding. 
    Other columns show best-fit expectation values and uncertainties for a free fit including a \SI{200}{\gevm} WIMP signal component. The best-fit signal cross-section is \bestfitfreesignalxsec. In addition to the expectation values in the full ROI, we include the expectation values in a signal-like cS1,cS2 region containing the $50\%$ of signal in  with the best signal-to-background ratio. This region is  indicated in Figure~\ref{fig:data_cs1cs2r} with an orange dashed contour. 
    The best-fit and pre-unblinding values agree within uncertainties for all components which include an ancillary constraint term.
    }
    \label{tab:backgrounds}
    \begin{tabular}{l |r | r | r}
    \hline
    \hline
         & \multicolumn{1}{c|}{Nominal} & \multicolumn{2}{c}{Best Fit} \\
    \hline
         & \multicolumn{2}{c|}{ROI} & Signal-like \\
    \hline
        {\er} & \nominaler  & \expectationerbestbothall& \expectationerbestbothsignallike \\
        Neutrons & \nominalradiogenictotal & \expectationradiogenictotalbestbothall & \expectationradiogenictotalbestbothsignallike\\
        \cevns &\expectationcevnstotalnominalbothall & \expectationcevnstotalbestbothall & \expectationcevnstotalbestbothsignallike\\
        AC & \expectationACtotalnominalbothall &  \expectationACtotalbestbothall & \expectationACtotalbestbothsignallike\\
        Surface & \expectationwallnominalbothall &  \expectationwallbestbothall & \expectationwallbestbothsignallike\\
    \hline
    %\hline
    Total Background  & 154 & \expectationbkgtotalbestbothall& \expectationbkgtotalbestbothsignallike \\
    WIMP  & - &\expectationsignalbestbothall & \expectationsignalbestbothsignallike \\
    \hline
    Observed & - & \neventstotal & \neventsreferenceregion \\
    \hline
    \hline
    \end{tabular}
\end{table}

The remaining contribution to the {\nr} background is predominately due to {\cevns} from \isotope[8]{B} solar neutrinos.  The rate is constrained by measurements of the \isotope[8]{B} flux~\cite{CEvNS_rate}, but the total uncertainty of the expectation value is dominated by the detector response model uncertainties. 
The number of cosmogenic neutrons is conservatively estimated to be less than \num{0.01} events after MV tagging~\cite{XENON1T:2014eqx}, not including the additional suppression by the NV. Thus, this background is considered to be negligible.

The {\er} background is dominated by $\upbeta$-decays of \isotope[214]{Pb} originating from the decay of \isotope[222]{Rn} in the LXe. Solar neutrino-electron scattering, \isotope[85]{Kr} and $\upgamma$-rays emitted by detector materials also contribute to the {\er} background~\cite{XENON:2022lower}. The {\er} response model fit was updated after unblinding of the main data to use the same data quality selections as of this study, compared to~\cite{XENON:2022lower}. 
Prior to unblinding, {\nominaler} events are found in the {\er} band of the ROI.

\begin{figure}[t]
    \centering
    \includegraphics[width=\columnwidth]{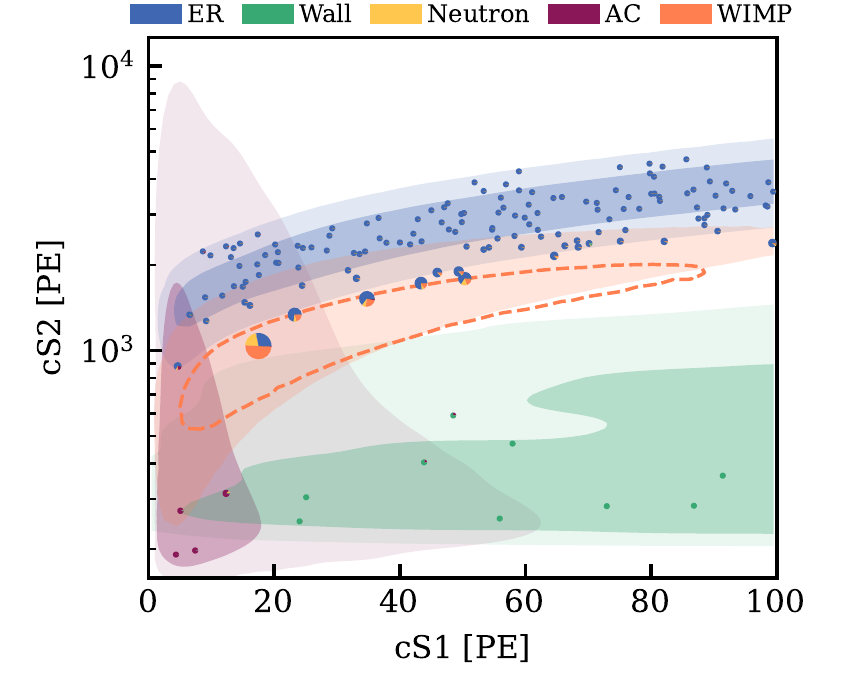}
    \caption{DM search data in the cS1-cS2 space. Each event is represented with a pie-chart, showing the fraction of the best-fit model, including the expected number of \SI{200}{\gevm} WIMPs (orange) evaluated at the position of the event. The size of the pie-charts is proportional to the signal model at that position. Background probability density distributions are shown as \SI{1}{\sigma} (dark) and \SI{2}{\sigma} (light) regions as indicated in the legend for {\er} (blue), AC (purple) and surface (green, \qmark{wall}). The neutron background (yellow in pies) has a similar distribution to the WIMP (orange filled area showing the \SI{2}{\sigma} region).
    The orange dashed contour contains a signal-like region which is constructed to contain 50\% of a \SI{200}{\gevm} WIMP signal with the highest possible signal-to-noise ratio.
    }
    \label{fig:data_cs1cs2r}
\end{figure}

Data-driven models are constructed for AC events and surface background events. The AC background is concentrated at low S1 and S2, and is therefore a particular challenge for low-mass WIMP searches. The model is constructed from a synthetic dataset made from isolated S1s and S2s using the method in~\cite{XENON:2020gfr}. Looser cuts in the near-wire region give a \tapprox$6$ times larger AC rate for this region compared to the rest of the TPC.  Background sidebands and \isotope[220]{Rn} and \isotope[37]{Ar} calibration data were used to validate the AC model, and the rate is estimated with an uncertainty of  better than \SI{5}{\%}.
The surface background model is constructed from \isotope[210]{Po} events originating from the TPC walls, using a similar method as in~\cite{xe1t_ana_paper_2}. The data is described in radius using a parametric likelihood fit based on events found below the blinded region. cS1 and cS2 are modeled using a kernel density estimation derived from events reconstructed outside of the TPC. 
The wall model is validated using the unblinded WIMP region outside of the FV as a sideband. The expected values for both backgrounds are summarized in Table \ref{tab:backgrounds} and their distributions in the (cS1, cS2) space are shown in Figure \ref{fig:data_cs1cs2r}. In addition, an extended version of Table \ref{tab:backgrounds} differentiating the near and far wire region can be found in Table \ref{tab:backgroundsextended} in the supplemental materials.

The statistical analysis of the WIMP search data uses toy MC simulations of the experiment to calibrate the distribution of a log-likelihood-ratio test statistic as in~\cite{xe1t_ana_paper_2,Baxter:2021pqo}. Four terms make up the likelihood: two search-data terms for events near and far from the transverse wires, an {\er} calibration term and a term representing ancillary measurements of parameters. The first three are extended unbinned likelihoods in cS1, cS2, as well as {\radius} for the first term. All three terms have the same form as equation (21) in~\cite{xe1t_ana_paper_2}. The two search-data likelihoods include components for the {\er}, AC, surface, {\cevns} and radiogenic neutron backgrounds, as well as the WIMP signal. The {\rnttz} calibration term includes the {\er} model as well as an AC component. The expected number of events for each component is a nuisance parameter in the likelihood.
In addition, two shape parameters for the {\er} model are included, and a parameter representing the uncertainty of the expected number of signal events given the {\nr} response model. The ER shape parameters mainly modify the signal-like ER tail below $S1=\SI{10}{\PE}$, where they allow the signal-like {\er} tail below the median S2 expected from a \SI{200}{\gevm} WIMP to vary between $0.009$ and $0.017$ at $60\%$ confidence level. The signal shape is fixed, as even a large signal excess would be small enough that the calibration constraints would dominate. The signal expectation value for a certain cross-section is included as a nuisance parameter. 
The ancillary measurement term includes Gaussians representing the measurements constraining the AC, radiogenic, surface and {\cevns} rates, and the uncertain signal expectation. 

The signal {\nr} spectrum is modeled with the Helm form factor for the nuclear cross section~\cite{lewin_review_1996}, and a standard halo model with parameters fixed to the recommendations of~\cite{Baxter:2021pqo}. The main change from previous XENON publications is an updated local standard of rest velocity of \SI{238}{\km\per\s}~\cite{bland-hawthorn_galaxy_2016,abuter_improved_2021}. 
The {\nr} model fit to calibration data is used to construct a model for the signal in cS1 and cS2.

\begin{figure}[t]
    \centering
    \includegraphics[width=\columnwidth]{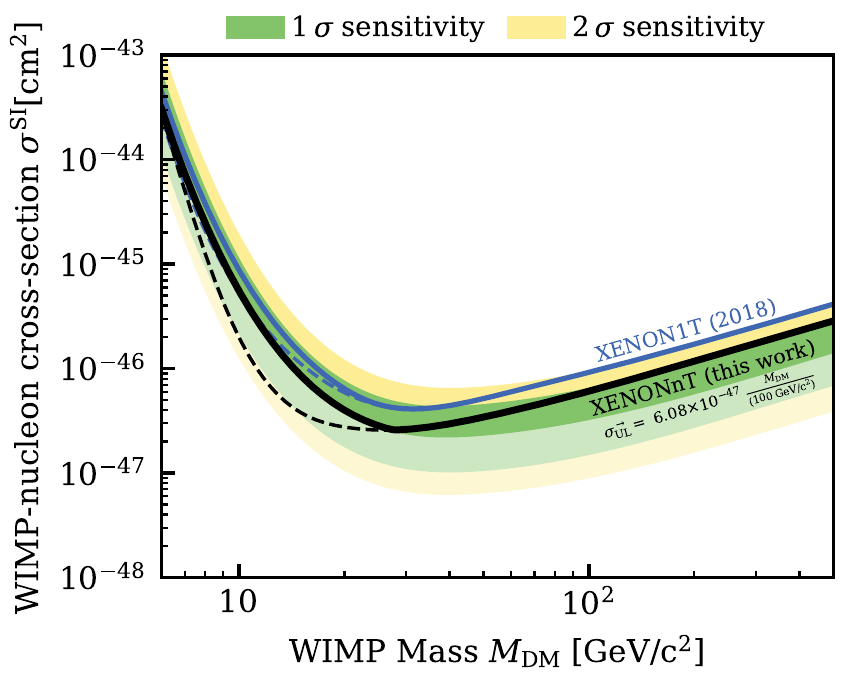}
    \caption{Upper limit on spin-independent WIMP-nucleon cross section at \SI{90}{\percent} confidence level (full black line) as a function of the WIMP mass. 
    A power-constraint is applied to the limit to restrict it at or above the median unconstrained upper limit.
    The dashed lines show the upper limit without a power-constraint applied. 
    The \SI{1}{\sigma} (green) and \SI{2}{\sigma} (yellow) sensitivity bands are shown as shaded regions, with lighter colors indicating the range of possible downwards fluctuations. 
    The result from XENON1T~\cite{XENON:2018voc} is shown in blue with the same power-constraint applied. At masses above $\tapprox\SI{100}{\gevm}$, the limit scales with mass as indicated with the extrapolation formula.
    }
    \label{fig:limit_si}
\end{figure}

After unblinding, the ROI contains {\neventstotal} events, {\neventsblinded} of which were in the blinded WIMP region. The data is shown in Figure~\ref{fig:data_cs1cs2r}, and the best-fit expectation values are in Table~\ref{tab:backgrounds}. The binned GOF test indicates no large-scale mismodelling ($\textrm{p}=\sciencebinnedgof$). 
At high cS1, $\gtrapprox \SI{50}{\PE}$, we observe more events which are consistent with ER events than our model or calibration data predicts, in particular between cS1s of \SI{50}{\PE} and \SI{75}{\PE}.
Of the {\neventsblinded} former blinded events, {\neventsupperrighthalf} are found in the upper right half of the horizontal event distribution, with no correlation with the transverse wires observed (see Figure~\ref{fig:data_xy}). The {\rnttz}, {\kretm} and {\argon} calibration datasets do not exhibit any asymmetry, nor is any seen in the acceptances evaluated in the $\mathrm{X},\mathrm{Y}$ plane for any of the applied cuts.

The WIMP discovery p-value indicates no significant excess ($\textrm{p}\geq\minimumpvalue$, with the minimum for masses above \tapprox$\SI{100}{\gevm}$), and the resulting limits on spin-independent interactions are shown in Figure~\ref{fig:limit_si}, with spin-dependent limits included in Figures~\ref{fig:limitsdp} and~\ref{fig:limitsdn} of the supplementary material. 
To constrain large downwards fluctuations, the limit is subjected to a power-constraint following~\cite{Cowan:2011an}. We choose a very conservative power threshold of $50\%$, higher than that advocated in~\cite{Baxter:2021pqo}, as that paper mistakenly defined the power-constraint in terms of discovery power when settling on a threshold of $15\%$. See the supplementary materials for further discussion. For spin-independent interactions the lowest upper limit is {\limitmassbest} at {\wimpmassbest} and 90\% confidence level (CL). At masses above \tapprox$\SI{100}{\gevm}$, the limit is {\highmasslimitfunction}. 
%To constrain large downwards fluctuations, the limit is subjected to a power-constraint following~\cite{Cowan:2011an}, using the \qmark{rejection power}, and a higher power-constraint of $50\%$ than recommended in~\cite{Baxter:2021pqo}, where the constraint is erroneously defined in terms of discovery power. See the supplementary materials for further discussion.
For spin-independent interactions the lowest upper limit is {\limitmassbest} at {\wimpmassbest} and 90\% CL. At masses above \tapprox$\SI{100}{\gevm}$, the limit is {\highmasslimitfunction}. 

In conclusion, a blind analysis of {\livetime} of science data with a total exposure of {\exposure} has been performed. 
The best fit to the data is compatible with the background-only hypothesis. 
The experiment achieved an {\er} background level of {\errateperkev}, $\tapprox 5$ times lower than XENON1T, with comparable detector resolutions, and energy threshold. 
This results in a sensitivity improvement with respect to XENON1T by a factor of $1.7$ at a WIMP mass of \SI{100}{\gevm}.

Currently, XENONnT continues to take data, with a further reduced \isotope[222]{Rn} {\er} background, using the radon distillation system with combined gaseous and liquid xenon flow.
Subsequent data-taking is planned with the NV operating as designed, with Gd-sulphate-octahydrate loaded into the water \cite{EGADS, SuperKamiokande} to increase the neutron tagging efficiency to $\tapprox$\SI{87}{\percent}  with a lower overall lifetime reduction~\cite{XENON:2020sensi}. 

We gratefully acknowledge support from the National Science Foundation, Swiss National Science Foundation, German Ministry for Education and Research, Max Planck Gesellschaft, Deutsche Forschungsgemeinschaft, Helmholtz Association, Dutch Research Council (NWO), Weizmann Institute of Science, Israeli Science Foundation, Binational Science Foundation, Fundacao para a Ciencia e a Tecnologia, R\'egion des Pays de la Loire, Knut and Alice Wallenberg Foundation, Kavli Foundation, JSPS Kakenhi and JST FOREST Program in Japan, Tsinghua University Initiative Scientific Research Program and Istituto Nazionale di Fisica Nucleare. This project has received funding/support from the European Union’s Horizon 2020 research and innovation programme under the Marie Sk\l{}odowska-Curie grant agreement No 860881-HIDDeN. Data processing is performed using infrastructures from the Open Science Grid, the European Grid Initiative and the Dutch national e-infrastructure with the support of SURF Cooperative. We are grateful to Laboratori Nazionali del Gran Sasso for hosting and supporting the XENON project.

% =====================================
\bibliography{bibliography.bib}

\setcounter{secnumdepth}{2}

\clearpage
\setcounter{figure}{0}   
\appendix
\renewcommand\thefigure{\thesection.\arabic{figure}}    
\section{Supplementary Material}
\subsection*{Limits on spin-dependent WIMP interactions}
In addition to spin-independent WIMP-nucleon interactions, WIMPs may couple to the nuclear spin. Following the procedure of~\cite{XENON:2019rxp}, and using the form factors of~\cite{Klos:2013rwa}, we place upper limits on this interaction assuming couplings to either only protons or only neutrons in the nucleus, $\sigma^\mathrm{SD}_{\chi p}$ and $\sigma^\mathrm{SD}_{\chi n}$, respectively. Unlike the spin-independent case, where detailed nuclear form factors fit the Helm form factor well, the form factors for spin-dependent interactions are subject to larger uncertainties~\cite{Hoferichter:2016nvd}. We consider a detailed exploration of the range of this uncertainty for spin-dependent responses beyond the scope of this work, as the uncertainty predominately changes the expected recoil rate, so that the results presented here can be re-scaled. However, we note that in e.g.~\cite{LZ:2022dm}, the range of recoil rates for these fluxes span approximately one order of magnitude.

\begin{figure}[htp]
    \centering
    \subfloat[\label{fig:limitsdp}]{%
    \includegraphics[width=0.99\columnwidth]{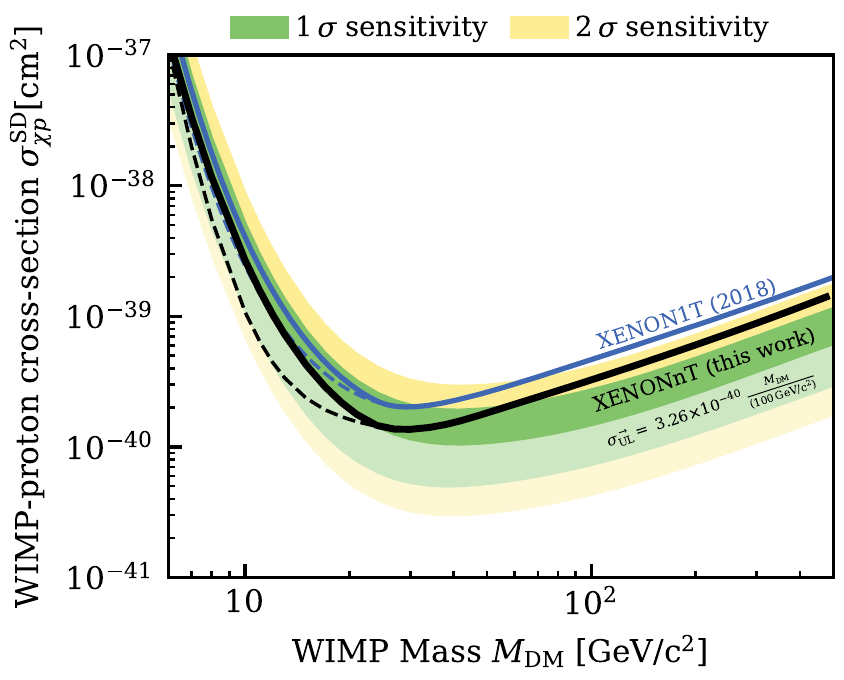}%
    }
    
    \subfloat[\label{fig:limitsdn}]{%
    \includegraphics[width=0.99\columnwidth]{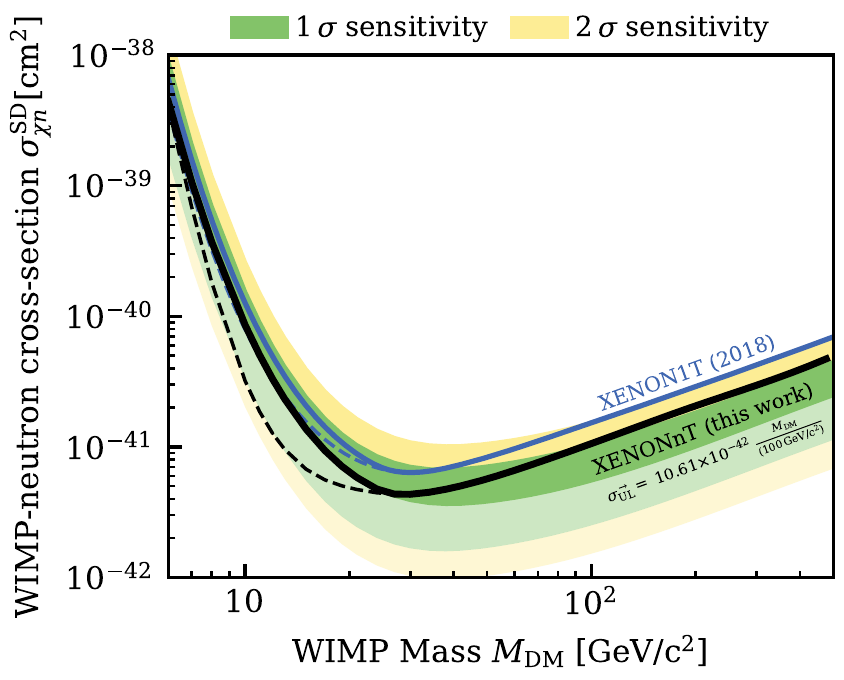}%
    }
    \caption{Upper limit on spin-dependent WIMP-proton (a) and WIMP-neutron (b) interactions at \SI{90}{\percent} confidence level (full black line) as a function of the WIMP mass. The limit is constrained by a power-constraint restricting upper limits to lie at or above the median unconstrained upper limit. The \SI{1}{\sigma} (green) and \SI{2}{\sigma} (yellow) sensitivity bands are shown as shaded regions, with lighter colors indicating the range of possible downwards fluctuations. The dashed line shows the upper limit without a power-constraint applied. Results from XENON1T~\cite{XENON:2019rxp} are shown in blue with the same power-constraint applied. At masses above $\tapprox\SI{100}{\gevm}$, the limit scales with mass as indicated with the extrapolation formula.
    \label{fig:limit_sd}}
\end{figure}

\clearpage
\subsection*{Spatial event distribution}
%try figure numbering from https://tex.stackexchange.com/questions/85776/change-figure-numbering-for-appendix
The distribution of events in the horizontal plane enters the likelihood only as radius {\radius} or in the near-wire/far-wire partition. Therefore, no dedicated GOF tests were defined in this plane prior to unblinding, and results are expected to be robust against a potential mismodelling that does not affect these variables.

\begin{figure}[htp]
    \centering
    \includegraphics[width=\columnwidth]{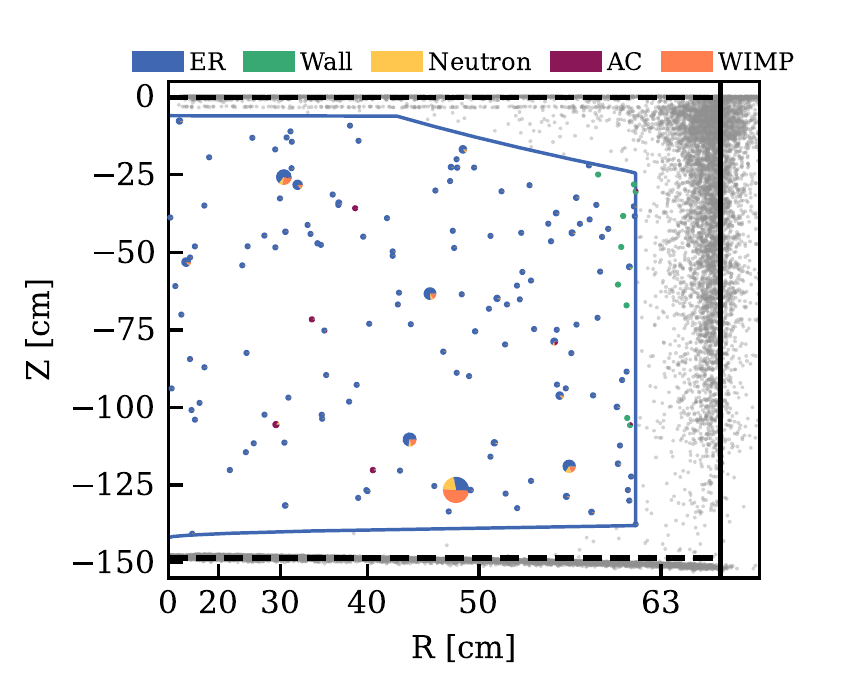}
    \caption{Spatial distribution of the search data in the {\fiducialvolume} fiducial volume (blue line). 
    Each event is represented with a pie-chart, showing the fraction of the best-fit PDF including a \SI{200}{\gevm} WIMP evaluated at the position of the event, color-coded as in Figure~\ref{fig:data_cs1cs2r}.  Events reconstructed outside of the fiducial volume are colored in gray. Black dashed lines depict the boundaries of the sensitive volume given by the cathode and gate positions. The TPC radius is indicated by a vertical black line.}
    \label{fig:data_rz}
\end{figure}

\begin{figure}[!h]
    \centering
    \includegraphics[width=\columnwidth]{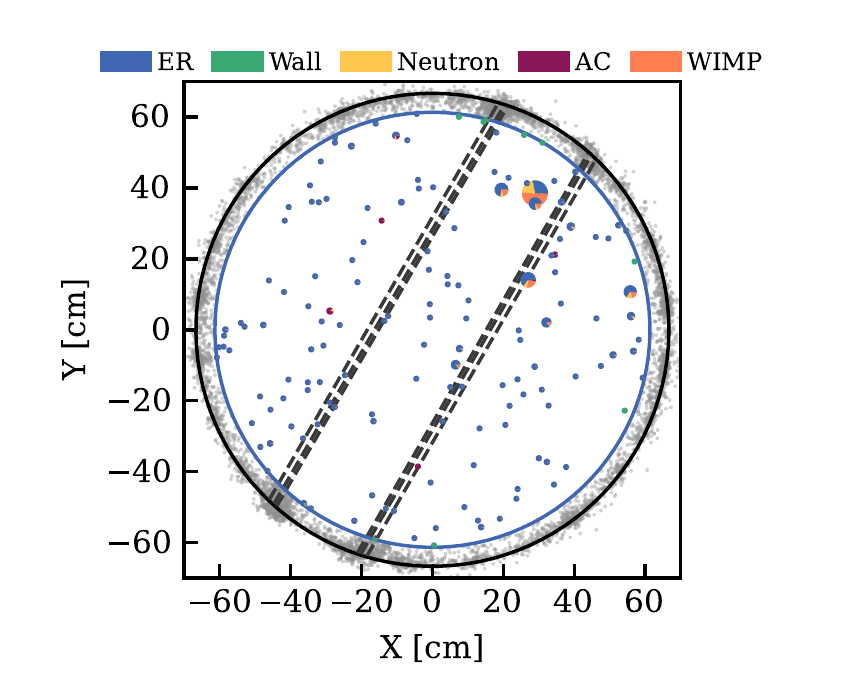}
    \caption{Distribution of the search data in the horizontal plane. 
    Each event is represented with a pie-chart, showing the fraction of the best-fit PDF including a \SI{200}{\gevm} WIMP evaluated at the position of the event, color-coded as in Figure~\ref{fig:data_cs1cs2r}. Events with a  reconstructed radius outside of the fiducial volume are shown in gray. 
    The inner (outer) circle indicates the FV selection (TPC radius), and dashed lines show the transverse wire positions.
    }
    \label{fig:data_xy}
\end{figure}

Figure \ref{fig:data_rz} and \ref{fig:data_xy} show the spatial distribution of all events found in the ROI after unblinding. The lower number of events next to the TPC wall for low z in Figure \ref{fig:data_rz} is due to a charge insensitive region caused by an inhomogeneity of the drift field. Field inhomogeneities also cause a variation in drift speed as function of $R$ leading to a small bias in the reconstructed z-position as it can be seen from the slight bending of the near cathode events in Figure \ref{fig:data_rz}. The bias is accounted for in the estimation of the fiducial volume and mass. The reconstruction bias is about \SI{+2}{\centi \meter} at the bottom outer edge of the fiducial volume and changes to roughly \SI{-0.4}{\centi \meter} at a radius of \SI{40}{\centi \meter}. 
 
Figure \ref{fig:data_xy} shows clusters with a higher density of events near the transverse wires as well as localised over densities in a periodic pattern close to the TPC wall outside of the fiducial volume. The former is caused by the higher rate of AC events and overall less strict requirements on the S2 width near the transverse wires as explained in the main body of the paper. The latter is an artefact due to the structure of the TPC PTFE cylinder which is composed of pillars and panels. The over densities are localised near the TPC pillars.

In total 13 of the 16 former blinded events are found in the upper right half of the horizontal event distribution. The events are neither found near the transverse wires nor the position of the single to few electron S2s burst which were localized in a \SI{10}{\centi \meter} radius around (\SI{5}{\centi \meter},\SI{-20}{\centi \meter}) in $X$ and $Y$. Additional test were carried out after unblinding to check if any systematic bias was overlooked during the development of corrections or data quality selections. The {\rnttz} {\er} band calibration data taken before SR0 was tested for a similar asymmetry for all calibration data points found in the lowest \SI{5}{\percent} of the {\er} band in cS2. Only a weak correlation with the transverse wires was found, as expected due to the \SI{10}{\percent} higher relative selection efficiency, but no asymmetry in the horizontal event distribution. The {\argon} calibration data taken after SR0 and {\kretm} taken every second week during SR0 were also tested for non-uniformity. In both cases no indication is found of an asymmetric bias in the event reconstruction. The impact of each data quality cut on the unblinded data was tested in several parameter spaces. The selections showed only expected correlations, e.g. the impact of the radius of the FV on the surface background, but none of the selections showed any behavior which could explain the observed asymmetry.


\clearpage
\subsection*{Comparison of upper limits}
\begin{figure}[ht]
    \centering
    \includegraphics[width=\columnwidth]{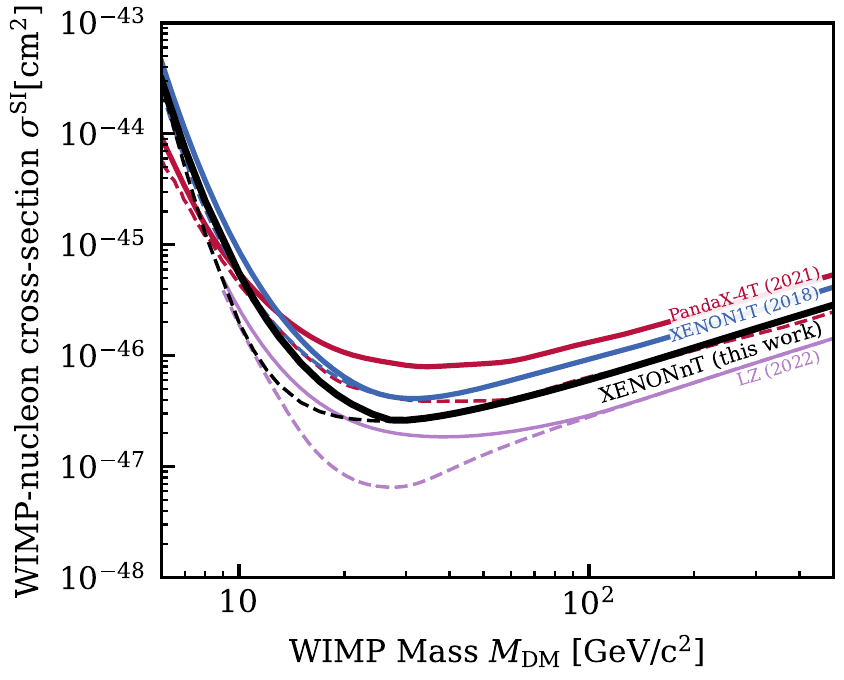}
    \caption{Upper limits on spin-independent WIMP-nucleon cross section at \SI{90}{\percent} confidence level for this work (black lines), LZ~\cite{LZ:2022dm} (purple lines, preprint),  PandaX-4T~\cite{PandaX:2021pan} (red lines) and XENON1T~\cite{XENON:2018voc} (blue lines). For PandaX and LZ, dashed lines represent their published result, for XENON results the dashed lines represent limits without a power-constraint applied. Full lines for each experiment represent a limit that is power-constrained to always lie at or above the median un-constrained limit. 
    }
    \label{fig:limit_allexperiments}
\end{figure}

Figure~\ref{fig:limit_allexperiments} compares this work to other recent results, both with and without a power-constraint applied consistent with the original PCL recommendation.
In order to not place limits on models for which an experiment has low sensitivity, a set of recommendations adopted by LXe dark matter experiments~\cite{Baxter:2021pqo} recommends using a power-constraint~\cite{Cowan:2011an}. The set of recommendations erroneously defines sensitivity in terms of discovery power, while it should be in terms of \qmark{rejection power}; the probability for a certain signal to be excluded given the no-signal hypothesis. This rejection power corresponds to the quantile of upper limits for that signal, as used to produce the conventional sensitivity bands.
The power-constrained limit is defined by setting a signal size threshold corresponding to a certain rejection power, and only placing upper limits at or above this threshold. 
This aims both to limit arbitrarily low limits being set by a systematic fluctuations, and moderates the effect on the upper limit of mis-modelling, in particular overestimated backgrounds. 
The choice of threshold rejection power is a fiducial one, and previous publications and the community recommendations (using discovery rather than rejection power) set it to correspond to the $\SI{-1}{\sigma}$ quantile of the limit distribution. 
Given the need to amend the recommendations, we choose a very conservative rejection power threshold of 0.5 for this work, corresponding to the median unconstrained limit. 

\begin{table*}[t]
    \centering
    \caption{Extended table from table~\ref{tab:backgrounds}, including expectation values computed separately for the wire region. }
    \label{tab:backgroundsextended}
    \begin{tabular}{l |r | r | r | r | r | r | r }
    \hline
    \hline
         & \multicolumn{1}{c|}{Nominal} & \multicolumn{6}{c|}{Best Fit} \\
    \hline
        & \multicolumn{4}{c|}{ROI} & \multicolumn{3}{c|}{Signal like} \\
    \hline
         & \multicolumn{2}{c|}{Full Volume} & off wire & on wire & Full Volume & off wire & on wire \\
    \hline
        {\er} & \nominaler  & \expectationerbestbothall& \expectationerbestoffall& \expectationerbestonall& \expectationerbestbothsignallike & \expectationerbestoffsignallike  & \expectationerbestonsignallike  \\
        Neutrons & \nominalradiogenictotal & \expectationradiogenictotalbestbothall &\expectationradiogenictotalbestoffall &\expectationradiogenictotalbestonall & \expectationradiogenictotalbestbothsignallike& \expectationradiogenictotalbestoffsignallike& \expectationradiogenictotalbestonsignallike\\
        \cevns &\expectationcevnstotalnominalbothall & \expectationcevnstotalbestbothall &\expectationcevnstotalbestoffall &\expectationcevnstotalbestonall & \expectationcevnstotalbestbothsignallike& \expectationcevnstotalbestoffsignallike& \expectationcevnstotalbestonsignallike\\
        AC & \expectationACtotalnominalbothall &  \expectationACtotalbestbothall & \expectationACtotalbestoffall & \expectationACtotalbestonall & \expectationACtotalbestbothsignallike& \expectationACnonwirebestoffsignallike& \expectationACwirebestonsignallike\\
        Surface & \expectationwallnominalbothall &  \expectationwallbestbothall & \expectationwallbestoffall & \expectationwallbestonall & \expectationwallbestbothsignallike& \expectationwallbestoffsignallike& \expectationwallbestonsignallike\\
    \hline
    %\hline
    Total Background  & 154 & \expectationbkgtotalbestbothall& \expectationbkgtotalbestoffall&\expectationbkgtotalbestonall&\expectationbkgtotalbestbothsignallike & \expectationbkgtotalbestoffsignallike & \expectationbkgtotalbestonsignallike \\
    WIMP  & - &\expectationsignalbestbothall &\expectationsignalbestoffall &\expectationsignalbestonall & \expectationsignalbestbothsignallike & \expectationsignalbestoffsignallike & \expectationsignalbestonsignallike \\
    \hline
    Observed & - & \neventstotal & \neventsnowire & \neventswire &  \neventsreferenceregion & \neventsreferenceregionnonwire & \neventsreferenceregionwire \\
    \hline
    \hline
    \end{tabular}
\end{table*}

\end{document}

%% file: expectation_definitions.tex
\newcommand{\expectationACnonwirebestoffsignallike}{ $0.16 \pm 0.04 $ }

\newcommand{\expectationACtotalbestbothall}{ $4.4_{ -0.8  }^{ +0.9 }$ }
\newcommand{\expectationACtotalnominalbothall}{ $4.3 \pm 0.9 $ }

\newcommand{\expectationACtotalbestbothsignallike}{ $0.32 \pm 0.06 $ }

\newcommand{\expectationACtotalbestoffall}{ $2.1 \pm 0.4 $ }

\newcommand{\expectationACtotalbestonall}{ $2.3_{ -0.4  }^{ +0.5 }$ }

\newcommand{\expectationACwirebestonsignallike}{ $0.16 \pm 0.04 $ }

\newcommand{\expectationbkgtotalbestbothall}{ $152 \pm 12 $ }

\newcommand{\expectationbkgtotalbestbothsignallike}{ $2.03_{ -0.15  }^{ +0.17 }$ }

\newcommand{\expectationbkgtotalbestoffall}{ $124_{ -9  }^{ +10 }$ }

\newcommand{\expectationbkgtotalbestoffsignallike}{ $1.61_{ -0.12  }^{ +0.13 }$ }

\newcommand{\expectationbkgtotalbestonall}{ $28 \pm 2 $ }

\newcommand{\expectationbkgtotalbestonsignallike}{ $0.42_{ -0.03  }^{ +0.04 }$ }

\newcommand{\expectationcevnstotalbestbothall}{ $0.23 \pm 0.06 $ }
\newcommand{\expectationcevnstotalnominalbothall}{ $0.23 \pm 0.06 $ }

\newcommand{\expectationcevnstotalbestbothsignallike}{ $0.022 \pm 0.006 $ }

\newcommand{\expectationcevnstotalbestoffall}{ $0.19 \pm 0.05 $ }

\newcommand{\expectationcevnstotalbestoffsignallike}{ $0.019 \pm 0.005 $ }

\newcommand{\expectationcevnstotalbestonall}{ $0.041 \pm 0.011 $ }

\newcommand{\expectationcevnstotalbestonsignallike}{ $0.0038 \pm 0.0010 $ }

\newcommand{\expectationerbestbothall}{ $135_{ -11  }^{ +12 }$ }

\newcommand{\expectationerbestbothsignallike}{ $0.92 \pm 0.08 $ }

\newcommand{\expectationerbestoffall}{ $110_{ -9  }^{ +10 }$ }

\newcommand{\expectationerbestoffsignallike}{ $0.76_{ -0.06  }^{ +0.07 }$ }

\newcommand{\expectationerbestonall}{ $24 \pm 2 $ }

\newcommand{\expectationerbestonsignallike}{ $0.167_{ -0.014  }^{ +0.015 }$ }

\newcommand{\expectationradiogenictotalbestbothall}{ $1.1 \pm 0.4 $ }

\newcommand{\expectationradiogenictotalbestbothsignallike}{ $0.42 \pm 0.16 $ }

\newcommand{\expectationradiogenictotalbestoffall}{ $0.9 \pm 0.4 $ }

\newcommand{\expectationradiogenictotalbestoffsignallike}{ $0.35 \pm 0.14 $ }

\newcommand{\expectationradiogenictotalbestonall}{ $0.18 \pm 0.07 $ }

\newcommand{\expectationradiogenictotalbestonsignallike}{ $0.07 \pm 0.03 $ }

\newcommand{\expectationsignalbestbothall}{ $2.6 $ }

\newcommand{\expectationsignalbestbothsignallike}{ $1.3 $ }

\newcommand{\expectationsignalbestoffall}{ $2.1 $ }

\newcommand{\expectationsignalbestoffsignallike}{ $1.1 $ }

\newcommand{\expectationsignalbestonall}{ $0.5 $ }

\newcommand{\expectationsignalbestonsignallike}{ $0.2 $ }

\newcommand{\expectationwallbestbothall}{ $12 \pm 2 $ }
\newcommand{\expectationwallnominalbothall}{ $14 \pm 3 $ }

\newcommand{\expectationwallbestbothsignallike}{ $0.35 \pm 0.07 $ }

\newcommand{\expectationwallbestoffall}{ $11 \pm 2 $ }

\newcommand{\expectationwallbestoffsignallike}{ $0.32 \pm 0.06 $ }

\newcommand{\expectationwallbestonall}{ $0.90_{ -0.17  }^{ +0.18 }$ }

\newcommand{\expectationwallbestonsignallike}{ $0.026 \pm 0.005 $ }

%% file: affiliations.tex
\newcommand{\bologna}{\affiliation{Department of Physics and Astronomy, University of Bologna and INFN-Bologna, 40126 Bologna, Italy}}
\newcommand{\chicago}{\affiliation{Department of Physics \& Kavli Institute for Cosmological Physics, University of Chicago, Chicago, IL 60637, USA}}
\newcommand{\coimbra}{\affiliation{LIBPhys, Department of Physics, University of Coimbra, 3004-516 Coimbra, Portugal}}
\newcommand{\columbia}{\affiliation{Physics Department, Columbia University, New York, NY 10027, USA}}
\newcommand{\lngs}{\affiliation{INFN-Laboratori Nazionali del Gran Sasso and Gran Sasso Science Institute, 67100 L'Aquila, Italy}}
\newcommand{\mainz}{\affiliation{Institut f\"ur Physik \& Exzellenzcluster PRISMA$^{+}$, Johannes Gutenberg-Universit\"at Mainz, 55099 Mainz, Germany}}
\newcommand{\heidelberg}{\affiliation{Max-Planck-Institut f\"ur Kernphysik, 69117 Heidelberg, Germany}}
\newcommand{\munster}{\affiliation{Institut f\"ur Kernphysik, Westf\"alische Wilhelms-Universit\"at M\"unster, 48149 M\"unster, Germany}}
\newcommand{\nikhef}{\affiliation{Nikhef and the University of Amsterdam, Science Park, 1098XG Amsterdam, Netherlands}}
\newcommand{\nyuad}{\affiliation{New York University Abu Dhabi - Center for Astro, Particle and Planetary Physics, Abu Dhabi, United Arab Emirates}}
\newcommand{\purdue}{\affiliation{Department of Physics and Astronomy, Purdue University, West Lafayette, IN 47907, USA}}
\newcommand{\rice}{\affiliation{Department of Physics and Astronomy, Rice University, Houston, TX 77005, USA}}
\newcommand{\stockholm}{\affiliation{Oskar Klein Centre, Department of Physics, Stockholm University, AlbaNova, Stockholm SE-10691, Sweden}}
\newcommand{\subatech}{\affiliation{SUBATECH, IMT Atlantique, CNRS/IN2P3, Universit\'e de Nantes, Nantes 44307, France}}
\newcommand{\torino}{\affiliation{INAF-Astrophysical Observatory of Torino, Department of Physics, University  of  Torino and  INFN-Torino,  10125  Torino,  Italy}}
\newcommand{\ucsd}{\affiliation{Department of Physics, University of California San Diego, La Jolla, CA 92093, USA}}
\newcommand{\wis}{\affiliation{Department of Particle Physics and Astrophysics, Weizmann Institute of Science, Rehovot 7610001, Israel}}
\newcommand{\zurich}{\affiliation{Physik-Institut, University of Z\"urich, 8057  Z\"urich, Switzerland}}
\newcommand{\paris}{\affiliation{LPNHE, Sorbonne Universit\'{e}, CNRS/IN2P3, 75005 Paris, France}}
\newcommand{\freiburg}{\affiliation{Physikalisches Institut, Universit\"at Freiburg, 79104 Freiburg, Germany}}
\newcommand{\napels}{\affiliation{Department of Physics ``Ettore Pancini'', University of Napoli and INFN-Napoli, 80126 Napoli, Italy}}
\newcommand{\nagoya}{\affiliation{Kobayashi-Maskawa Institute for the Origin of Particles and the Universe, and Institute for Space-Earth Environmental Research, Nagoya University, Furo-cho, Chikusa-ku, Nagoya, Aichi 464-8602, Japan}}
\newcommand{\laquila}{\affiliation{Department of Physics and Chemistry, University of L'Aquila, 67100 L'Aquila, Italy}}
\newcommand{\tokyo}{\affiliation{Kamioka Observatory, Institute for Cosmic Ray Research, and Kavli Institute for the Physics and Mathematics of the Universe (WPI), University of Tokyo, Higashi-Mozumi, Kamioka, Hida, Gifu 506-1205, Japan}}
\newcommand{\kobe}{\affiliation{Department of Physics, Kobe University, Kobe, Hyogo 657-8501, Japan}}
\newcommand{\ucla}{\affiliation{Physics \& Astronomy Department, University of California, Los Angeles, CA 90095, USA}}
\newcommand{\kit}{\affiliation{Institute for Astroparticle Physics, Karlsruhe Institute of Technology, 76021 Karlsruhe, Germany}}
\newcommand{\tsinghua}{\affiliation{Department of Physics \& Center for High Energy Physics, Tsinghua University, Beijing 100084, China}}
\newcommand{\ferrara}{\affiliation{INFN - Ferrara and Dip. di Fisica e Scienze della Terra, Universit\`a di Ferrara, 44122 Ferrara, Italy}}
\newcommand{\alsoatcoimbrapoli}{\affiliation{Coimbra Polytechnic - ISEC, 3030-199 Coimbra, Portugal}}
\newcommand{\alsoatuniheidelberg}{\affiliation{Physikalisches Institut, Universit\"at Heidelberg, Heidelberg, Germany}}
\newcommand{\alsoatroma}{\affiliation{INFN - Roma Tre, 00146 Roma, Italy}}

%% file: author.tex
\author{E.~Aprile}\columbia
\author{K.~Abe}\tokyo
\author{F.~Agostini}\bologna
\author{S.~Ahmed Maouloud}\paris
\author{L.~Althueser}\email{l.althueser@uni-muenster.de}\munster
\author{B.~Andrieu}\paris
\author{E.~Angelino}\torino
\author{J.~R.~Angevaare}\nikhef
\author{V.~C.~Antochi}\stockholm
\author{D.~Ant\'on Martin}\chicago
% ...
